# Hypothesis: Is percolative clustering an emerging paradigm in protein-protein interactions ?

**RV Krishnan**

Department of cellular and structural biology, University of Texas Health Science Center, 7703 Floyd Curl Drive, San Antonio, TX 78229, USA

Tel: (+1)-210-567-4571

Fax:(+1)-210-567-3803

Email: krsna@uthscsa.edu (corresponding author)

## Abstract

**Background**

The evolution, regulation and sustenance of biological complexity is determined by protein-protein interaction network that is filled with dynamic events. Recent experimental evidences point out that clustering of proteins has a vital role in many cellular processes.  Upsurge in fluorescence imaging methods has given a new spin to our ability to probe protein interactions in cellular and sub-cellular compartments. Despite the increasing detection sensitivity, quantitative information that can be obtained from these imaging methods is limited.  This is primarily due to (i) the difficulty in tracking the problem analytically and (ii) limitations in spatio-temporal resolution that can be achieved in interrogating living cells in real time.

**Hypothesis**

A novel point of view based on diffusion-driven percolative clustering is proposed here that can plausibly shed more light on the complex issues of protein-protein interactions. Since this model is open to computational analysis, it is quantitative in its premise. Besides being able to analyze the phenomenon, the power of any model is gauged by its ability to predict interesting and novel features of the phenomenon itself, which can subsequently be tested by additional experiments. To this end, an experimental assay based on fluorescence lifetime imaging is proposed to verify the validity of the percolation model.

**Implications**

Modeling the temporal evolution of the cluster distribution, bound clusters amidst unbound receptors, can give useful information on diffusion behavior of the proteins as well as on the binding constants of the complexes. Emerging trends in clustering algorithms based on the "mutual connectivity" of gene expression profiles are vital in the Human Genome sequencing era. Concepts in the collective dynamics of networks (e.g., "small worlds" network) hypothesized in various other circumstances may also be a plausible way of understanding biological networks. As a practical note, this model could be easily translated to a real time FRET image analysis module for extracting quantitative information about individual proteins and molecular assemblies in their native physiological environment. Furthermore, with careful experimental design, it is possible to extend the scope of this model to protein-lipid and lipid-lipid interactions as well.

# Background

Interactions among proteins in living cells can be dissected into two major events: first, the diffusive transport in cellular environments to explore the corresponding partners and second, interacting with the partner in a very specific manner (e.g antibody-antigen specificity). Understanding these two steps entails sensitive experimental assays to measure these specific interactions with high spatial/temporal resolution as well as quantitative methods to extract parameters that characterize the interactions. This commentary aims to present a hypothesis based on a percolative clustering network model that captures the essence of protein-protein interactions in living cells. Since this model is open to computational analysis, it is quantitative in its premise. Besides being able to analyze the phenomenon, the power of any model is gauged by its ability to predict interesting and novel features of the phenomenon itself, which can subsequently be tested by additional experiments. To this end, an experimental assay based on fluorescence lifetime imaging is proposed to verify the validity of the percolation model.

# Hypothesis and Discussion

**Invasion percolation model**

The conceptual basis for percolation is intuitively simple and has been used in a wide variety of fields such as mineralogy, forestry, polymer physics, porous media and granular transport [1,2]. Figure 1 depicts a cartoon of percolation concept. In a random network, percolation deals with connectivity among clusters that have some defined similarity with respect to the measured phenomenon. An intuitive percolation picture of interacting proteins can be easily understood as a two-state model where the individual states could represent, for example, bound and unbound ligand, or monomer and dimer formation in receptor population or different conformational states of the same protein. In the simplest case of biological organization, if we model the cell membrane as an infinite two-dimensional surface with a random distribution of lipids and proteins, every occupied node in Figure 1 can be thought of as a protein under investigation. Invasion percolation is the process by which a molecule moves or 'invades' through the lattice by any of several different transport processes (for e.g., ligands moving through the surface in search of their corresponding receptors to bind). An important feature of invasion percolation is that it is inherently dynamic, allowing mobility for the molecule. As the ligands explore the receptor distribution for stable binding sites, their mobility is essentially governed by discrete steps in a "random-walk" – either due to normal or anomalous diffusion. Such a random-walk eventually leads to receptor-ligand binding as determined by parameters such as the diffusion constant, binding and dissociation rate constants, membrane viscosity. However biologically significant interactions are governed by just not a single pair of proteins but by a cooperative network of interacting proteins. Such cooperative interactions may be homotypic or heterotypic, eventually leading to local clustering of proteins. It is hypothesized here that percolative clustering of interacting proteins is a necessary step in protein-protein interactions and that it is the generalized paradigm in biological networks. The following paragraph explains in detail this hypothesis of percolative clustering in cellular context.

**Percolative clustering in protein interaction networks** : *a hypothesis*

The simplest mechanistic picture of protein movements in cellular environments is given by the classical Brownian motion which describes diffusive transport in fluids. According to this picture, proteins execute discrete random-walks and display interactions that are characteristic of interacting proteins. Inter- and intra-cellular communication is strongly influenced by these pair-specific interactions to create an apparent order in a seemingly random network. Immune response, wound healing, ion transport across the membranes, neuronal signaling, regulation of gene expression and cell cycles are some of the classic examples of cellular efforts to create an 'order out of chaos'. A common denominator of these interaction networks is that there is a strong local clustering of interacting proteins whose distribution and mobility determine the function of any protein pathway. Current wisdom from protein interaction pathways points out that these networks are highly interconnected. In spite of the fact that these interaction networks constantly change their configurations by the addition and removal of links in their local environment, there is a well-defined distribution of cluster sizes at any given time. Localized protein interactions determine the direction of the growth and stability of these clusters which in turn, manifest as a global response from the interaction network. Percolative clustering hypothesis therefore can provide a quantitative description of protein-protein interactions, the applicability of which is justified by the following reasons: (i) the premise of percolation is a random network model which describes well the protein interactions in cellular environment; (ii) percolation deals with the properties of clusters and mutual connectivity between them and this scenario has an immediate correspondence with the biologically relevant local clustering of interacting proteins; (iii) an analytical solution to the problem of protein interactions ( many-body problem) is very difficult owing to the fact that the initial state of the system can not be completely defined and percolation simulation can therefore describe complex interactions in real time with great redundancy in mutual connectivity and finally (iv) predictions from percolation model can be easily verified by fluorescence-based assays as described below. Evidence from various cellular phenomena such as endocytosis, signal transduction and cell-cell communication strongly suggest that percolative clustering may be a rule in biological interactions rather than just an exception. There have been a few reports on the application of percolation phenomenon in biological context, mainly in structural transegrity, cytoskeletal organization and in gene expression profiling [3,4]. Despite the simple premise and extensive computational developments in percolation, it is surprising that there are no serious efforts in the past to apply percolation modeling to understanding protein-protein interactions.

**Experimental approaches to probe percolative clusters in living cells**

The hypothesis of percolative clustering and its implications in understanding protein-protein interactions can be experimentally verified by biochemical (immunoprecipitation, protein affinity chromatography, crosslinking) and/or biophysical (diffusion measurements) techniques. It should be clearly understood however, that the information content obtained from various experiments may not be the same owing to the fact that every method probes different length- and time-scales of protein-protein interaction networks [5]. For instance, the size of protein clusters that can be measured by immunoprecipitation or crosslinking will be quite larger than that measured by sensitive diffusion measurements. Thus depending on the sensitivity of the experimental assay, one can obtain information from protein clusters which are microscopic to nanoscopic in dimension. In recent times, fluorescence imaging has revolutionized our ability to interrogate living cells. A striking similarity between percolation modeling and fluorescence imaging can be identified in the specificity and sensitivity of these approaches. For instance, in a percolation model, one traces the response of a single component amidst a plethora of interacting components. Similarly in fluorescence imaging, an antibody or protein is tagged with a fluorophore and its individual response amidst a variety of other proteins is being tracked. This makes percolation model an excellent numerical analog of fluorescence imaging. In the simple case of receptor-ligand interactions, a fluorescently labeled ligand serves as the tracer (Figure 1d). At lower concentrations of the ligand, the lattice contains very few clusters of bound receptors in an ocean of unbound receptors. Beyond a certain threshold concentration of ligand, clusters interconnect and can form an "infinite" or "spanning" cluster of bound receptors that may be a required step to initiate a biological signal. Experimentally, the mobility of a ligand can be monitored by diffusion methods such as single particle tracking, fluorescence correlation spectroscopy, fluorescence recovery after photobleaching[6,7]. Nevertheless, these methods are unable to provide high spatial resolution. Alternately, one can fluorescently label the receptor on the membrane and monitor the fluorescence anisotropy of the receptor in its bound/unbound states. Anisotropy reveals information on rotational correlation and hence depends on the shape and size of the molecular complex. In the two-state percolation representation, a bound receptor yields high anisotropy and unbound receptor yields a low anisotropy. In a recent excellent article, Mayor and colleagues reported on the nanoscale (< 5nm) organization of GPI-anchored proteins in living cell membranes by combining anisotropy measurements with theoretical modelling [8]. It is intriguing to note that percolative clustering model proposed here is in principle a generalized approach for understanding protein-protein interactions regardless of the size of the clusters as well as understanding steady-state and transient cellular processes.

**2D percolation clustering in membranes : an experimental demonstration**

Figure 2 exemplifies the applicability of percolation hypothesis in the study of receptor-receptor interactions mediated by a diffusing species of ligand (e.g., the classical case of hormone induced receptor dimerization). Among the various fluorescence assays that are being employed in the study of protein-protein interactions, Fluorescence Resonance Energy Transfer (FRET) microscopy has been very popular owing to its ability to detect inter- and intra- molecular interactions in the 1-10 nm range [9,10]. The difficulty in extracting molecular information from intensity based FRET approaches is primarily due to the highly nonlinear spectral profiles as well as to the artifacts involved in the measurements. Alternatively one can employ fluorescence lifetime imaging microscopy (FLIM) as a reliable assay for studying protein-protein interactions [11,12]. As fluorescence lifetime ($\tau$) is an intrinsic property of the fluorophore, it can be a reliable reporter of changes in the radiative decay rate induced by local environmental effects such as pH and ionic concentrations as well as energy transfer events. Interestingly, this situation mimics the clustering hypothesis that relies heavily on the influences of nearby neighbors. Figure 2 shows a modified percolation lattice which contains a mixture of receptors labeled with two different fluorophores used as a donor/acceptor FRET pair. There is an energy transfer event that occurs when the hormone brings neighbor receptors in proximity so that they are dimerized. FRET manifests as a reduction in donor lifetime which brings the percolation lattice again to the two-state representation. FRET is indicated by lower donor lifetime whereas a higher donor lifetime denotes no FRET. Top panel of Figure 2 shows an experimental verification of this prediction. Representative lifetime images of cells expressing cytokine receptor tagged with fluorescent proteins are given. Hormone application has been shown to induce receptor-receptor interactions [13]. The time evolution of this interaction scenario can be seen from the images as a function of duration of hormone stimulus. A percolative snapshot of this hormone induced receptor dimerization is shown in the lower panel of Figure 2. The system evolves from its initial state of globally higher donor lifetime to a state where the system as a whole has a large number of receptors with reduced lifetimes. This implies an "infinite" or "spanning" cluster of FRET pairs has formed (figure 2e). This global connectivity can be visualized by imaging individual cells/tissues and FRET can be visualized as rising from the dynamic modification of the percolation lattice. One should note that as the ligands execute diffusion-driven random walks amidst the receptor population, there exists a large number of identical receptor distributions for any given probability of being bound. Therefore, the observed binding is a net result of a stochastic exploration of different binding configurations and the concomitant energy minimization for the receptor-ligand complex. The necessity of several attempts prior to a successful occupation is reminiscent of noise reduction in diffusion-limited aggregation [14]. This noise reduction has the general effect of smoothing and leading to clusters that grow slowly but regularly. Modeling the temporal evolution of the cluster distribution, bound clusters amidst unbound receptors, can give useful information on diffusion behavior of the ligand as well as on the binding constants of the complex.

# Conclusions and perspectives

Although examples presented in this commentary have been restricted to interactions between two proteins in cell membrane it is readily possible to extend this approach to more complex, multiple-protein interactions by means of "correlated percolation" and also three-dimensional interactions.  Similarly, improved percolation models incorporating anomalous diffusion, which can arise due to obstacles and binding, can be used to more accurately represent diffusion kinetics of protein species [15,16].  It is interesting to observe that, owing to its intrinsic cluster structure, the percolation model can naturally incorporate the more recent hypotheses based on lateral heterogeneities in cell membranes such as 'domain hopping'  and "lipid raft" hypotheses [17,18]   Emerging trends in clustering algorithms based on the "mutual connectivity" of gene expression profiles are  vital in the Human Genome sequencing era [19].  Concepts in the collective dynamics of networks (e.g., "small worlds" network) hypothesized in various other circumstances may also be a plausible way of understanding biological networks [20].  As a practical note, this model could be easily translated to a real time FRET image analysis module for extracting quantitative information about individual proteins and molecular assemblies in their native physiological environment.  Furthermore, with careful experimental design, it is possible to extend the scope of this model to protein-lipid and lipid-lipid interactions as well.  Considering the rapid developments in laser technology, fast detectors, imaging probes and computational facilities, it is an exciting time to begin exploring new ways of evaluating biological organization.

## Authors' contributions

RVK conceived the hypothesis, carried out the lifetime imaging and drafted the manuscript.

## Acknowledgements

I acknowledge Prof. Brian Herman for his guidance and mentorship, Ms.Eva Biener and Prof. Arieh Gertler for cells and Drs. James Lechleiter, Meera Patturajan and Bijoy Thattaliyath for their valuable comments.

# Figures

**Figure 1: Schematic of percolation concept** (a-c) Consider a two-dimensional lattice of size (N x N) where every node can be occupied (filled) or empty (open). There is an equal probability (p) for every node to be occupied. If we assign randomly a certain number of nodes to be occupied at any point of time, there will be pN occupied nodes and (1-p)N empty nodes as determined by stochastic "random walk" steps executed by the occupant of the node. Adjacent nodes that are occupied are called *neighbors* and groups of neighbors are called *clusters*. Percolation mainly deals with the properties of these clusters (size, density etc,) and how connected these clusters are. Rules for assigning neighbor status are governed by the geometry ( square, triangular, honeycomb etc.) and the dimensionality of the lattice. Regardless of these rules, for a given distribution of clusters, a cluster density can be defined : $d_s$ = (number of occupied nodes/total number of nodes). Note that there is no inherent pattern to the distribution of the occupied nodes (i.e there is no initial bias) and there can be many different lattice configurations possible for a given probability of being occupied. Another parameter that defines the percolation phenomenon is percolation threshold, $p_c$ at which value individual clusters coalesce to form atleast one "infinite" or "spanning" cluster that provides connectivity from one end of the lattice to the other (b). Beyond $p_c$ there is atleast one spanning cluster and there may arise several alternate spanning clusters (c). Values of $p_c$ for square, honeycomb and cubic lattices are 0.5,0.65 and 0.25 respectively. The above schematic is just a snap-shot at any given time. In time-dependent situations (binding/unbinding transitions) the measured macroscopic properties depend sensitively on the timescales of measurement. (d) Cluster formation in a percolative network of receptor-ligand interactions.

**Figure 2 A percolation perspective of receptor-receptor interaction.** In this bond percolation problem, a distance $r_{ij}$ is defined between all pairs of points i and j such that small distances correspond to higher FRET efficiency [$E \propto (1/r_{ij}^6)$] and large distances to lower FRET. Probability of connecting two points $P(r_{ij}) = \exp(-r_{ij}^2/r_o^2)$ which varies continuously from 0 and 1. In this way, two points (i,j) are directed connected only if their distance $r_{ij} \ll r_o$ (a threshold distance dependent on the pair's spectral features and orientational degrees of freedom). (a-c) Fluorescence lifetime images of human embryonic kidney cells co-expressing prolactin receptors tagged with cyan-(PRLR-CFP) and yellow-(PRLR-YFP) fluorescent proteins (Image courtesy : Ms.Eva Biener and Prof. Arieh Gertler, Hebrew University, Rehovot, Israel). Cells were stimulated (0-30 minutes) with prolactin, a hormone which is known to induce interaction between prolactin receptors. Receptor dimerization kinetics was monitored in fixed specimens by fluorescence lifetime imaging microscopy (Reference 12). Decreased proximity between receptors is evidenced by energy transfer (FRET) between CFP and YFP which is expected to manifest as a reduction in CFP (donor) lifetime. Figure 2b shows a clear reduction in CFP lifetime at ~3 minutes which can be interpreted as the most probable distribution of dimerized receptors in these cells. (d-f) Percolative snap shots of this dimerization kinetics. Information on diffusion kinetics of hormone, hormone/receptor binding kinetics and receptor dimerization kinetics can be obtained by analyzing the time evolution of these clusters with a diffusion-driven invasion percolation model. Such a model can possess ability to predict biophysical response from homologous receptor families

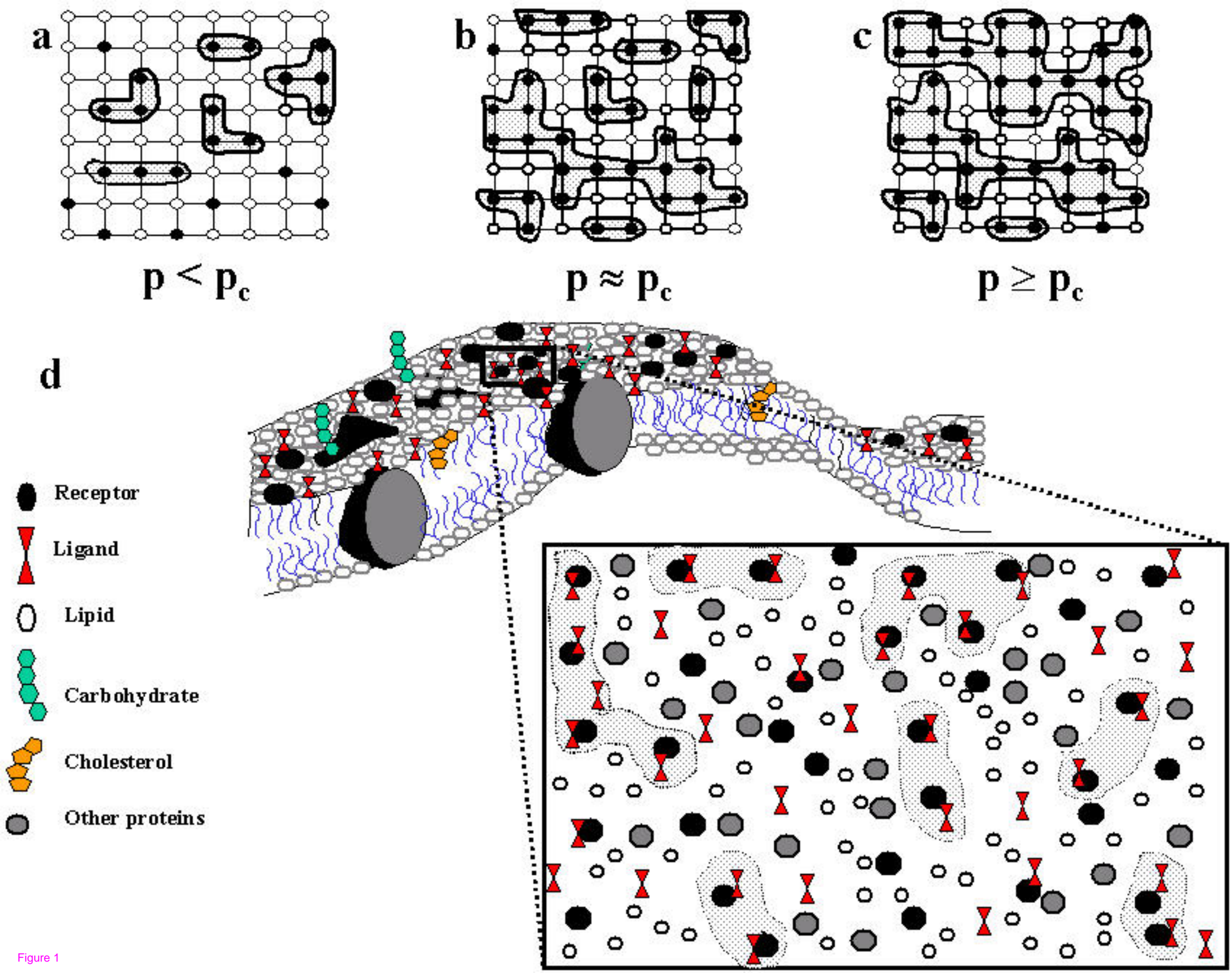

Figure 1

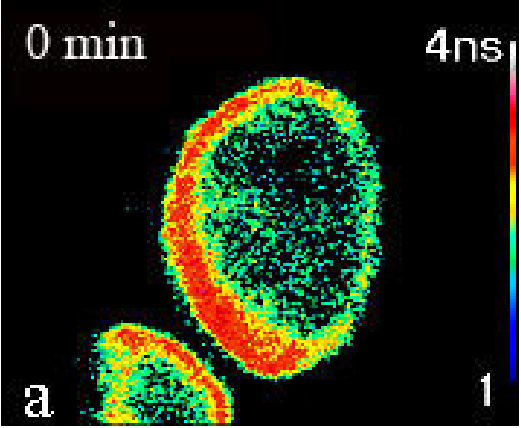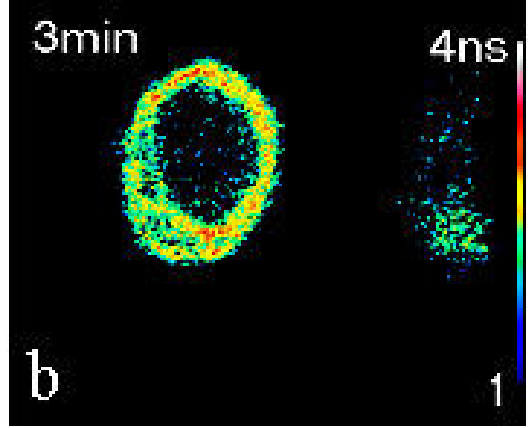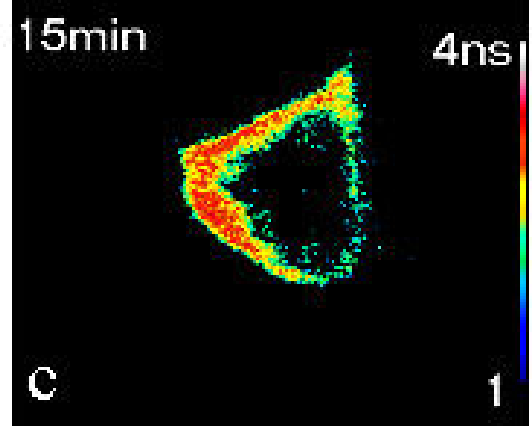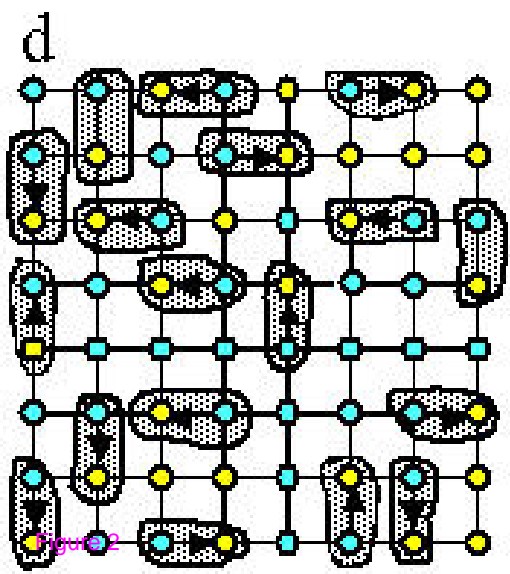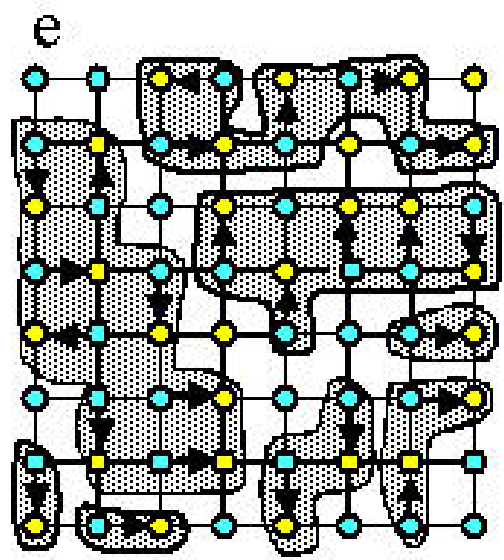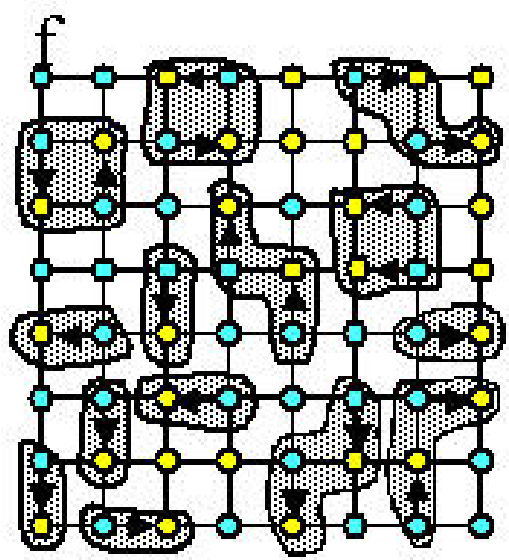